\documentclass[preprint,12pt]{aastex}
\usepackage{epsfig, amsmath, amsfonts}

\newcommand\bb[1] {   \mbox{\boldmath{$#1$}}  }

\newcommand\del{\bb{\nabla}}

\def\gtsima{$\; \buildrel > \over \sim \;$}
\def\gtsim{\lower.5ex\hbox{\gtsima}}

\newcommand\bcdot{\bb{\cdot}}
\newcommand\btimes{\bb{\times}}
\newcommand\vv{\bb{v}}

\newcommand\dd{\partial}
\newcommand\simga{\sigma}

\def\beq{ \begin{equation} }
\def\eeq{ \end{equation} }

\begin{document}


\title{\bf \Large
Global Model of Differential Rotation in the Sun}

\author{ Steven A. Balbus\altaffilmark{1,2, 4},
Henrik Latter\altaffilmark{1,3}, Nigel Weiss\altaffilmark{3}}

\altaffiltext{1}{Laboratoire de Radioastronomie, \'Ecole Normale
Sup\'erieure, 24 rue Lhomond, 75231 Paris CEDEX 05, France
  \texttt{steven.balbus@lra.ens.fr}}


\altaffiltext{2}{Institut universitaire de France,
Maison des Universit\'es, 103 blvd.\ Saint-Michel, 75005
Paris, France}

\altaffiltext{3}{DAMTP, Wilberforce Road, Cambridge CB3 0HA, UK}

\altaffiltext{4}{Princeton University Observatory, Peyton Hall, Princeton University, Princeton NJ 08540}

\begin{abstract}
The isorotation contours of the solar convective zone (SCZ) show three
distinct morphologies, corresponding to two boundary layers (inner and
outer), and the bulk of the interior.  Previous work has shown that the
thermal wind equation together with informal arguments on the nature
of convection in a rotating fluid could be used to deduce the shape of
the isorotation surfaces in the bulk of the SCZ with great fidelity,
and that the tachocline contours could also be described by relatively
simple phenomenology.  In this paper, we show that the form of these
surfaces can be understood more broadly 
as a mathematical consequence of the thermal wind equation and a narrow
convective shell.  The analysis does not yield the angular velocity
function directly, an additional surface boundary condition is required.
But much can already be deduced
without constructing the entire rotation profile.  
The mathematics may be combined with dynamical
arguments put forth in previous works to the mutual benefit of each.
An important element of our approach is to regard the constant angular
velocity surfaces as an independent coordinate variable for what is
termed the ``residual entropy,''  a quantity that plays a key role in the
equation of thermal wind balance.  The difference between the dynamics of
the bulk of the SCZ and the tachocline is due to a different functional
form of the residual entropy in each region.  We develop a unified theory
for the rotational behavior of both the SCZ and the tachocline, using
the solutions for the characteristics of the thermal wind equation.
These characteristics are identical to the isorotation contours in
the bulk of the SCZ, but the two deviate in the tachocline.  The outer
layer may be treated, at least descriptively, by similar mathematical
techniques, but this region probably does not obey thermal wind balance.

\end{abstract}

\maketitle

\section {Introduction}
Helioseismology has revealed a global pattern of differential
rotation in the solar convective zone (SCZ) comprising three distinct,
readily apparent, regions.
In the bulk of the SCZ (region 1), the isorotation contours appear mainly
as quasi-radial spokes in meridional planes
(corresponding in three dimensions to cones).  The contours
change relatively little and only very gradually near the poles
and equator, but in the boundary between the SCZ and the uniformly rotating
radiative interior (the ``tachocline''), 
the isorotation contours are abruptly wrenched into
quasi-spherical shells (region 2).  Finally, in the outermost few
percent of the Sun's exterior layers, the SCZ isorotation spokes
gradually acquire lengthening tangential tails stretching
from the parent spoke toward the equator (region 3).  The challenge
for the theorist is to account for this global organization as
economically as possible, a task that motivates the current paper.

Despite the intrinsic dynamical complexity of a rotating, convecting
fluid there are some indications that the solar rotation profile
may obey rather simple mathematical laws.  In particular, much
of the convective zone may be governed by a reduced form of the vorticity
equation, known as the thermal wind equation, or TWE 
(Thompson et al. 2003).  The TWE involves a balance between 
inertial angular velocity gradients and baroclinic driving, the latter
arising from angular entropy gradients that may be present.  

The TWE is not, by itself, sufficient to solve for the rotation profile,
since it is a single equation containing two unknowns, the angular velocity
$\Omega$ and entropy $S$.  
It has been suggested, however, that $S$ and $\Omega$
may be {\em a priori} functionally related, and that this relationship
could be exploited to solve the TWE. For example, Balbus (2009) pointed out
that a certain class of magnetized baroclinic modes were marginally
stable if constant $S$ and $\Omega$ surfaces coincided.  With $S$ a
function of $\Omega$ only, the TWE can be formally,
but quite usefully, solved.  The solution does not yield 
the entire spatial structure of $\Omega$, rather it yields a parameterized
form
for the surfaces on which $\Omega$ is a constant (see equation [3], below).
The fit to the data is strikingly good.

In a subsequent analysis, Balbus et al. (2009; hereafter
BBLW) argued that magnetic fields were not essential to establishing
the $S-\Omega$ functional relationship.
Under conditions expected in the SCZ, the effect of
differential rotation on thermally convecting cells will be to distort  
surfaces of constant entropy into surfaces of constant angular velocity
$\Omega$.  More precisely, one expects the surfaces of constant {\em residual}
entropy---the portion of the entropy in excess of the destabilizing
radial entropy profile needed to drive convection---to tend to
coincide with surfaces of constant angular velocity.  In the TWE, a
functional relation between residual entropy $S'$ and $\Omega$ works
just as well as an $S-\Omega$ functional relationship, since it is only
the angular derivatives of $S$ that enter into the 
equation: the difference between $S$ and $S'$
is a function only of radius $r$.  But the distinction
between residual and total entropy is of course otherwise important.   
In numerical simulations
of the SCZ that most closely resemble the Sun, for example, surfaces of constant $S$   
are nearly spherical shells, whereas surfaces
of constant $S'$ indeed correspond much more closely to 
those of constant angular velocity (Miesch, Brun, \&
Toomre 2006).

The mathematical implementation of the constraint of
shared surfaces of constant residual entropy and
rotation may be written
\beq\label{one}
S(r, \theta) = f(\Omega^2) +S_r(r)
\eeq
where $S(r,\theta)$ is the entropy, $r$ and $\theta$ are the standard
spherical radius and colatitude angle, $\Omega$ is the angular velocity,
$f$ is an arbitrary function of $\Omega^2$, and $S_r$ is the spherical
entropy profile needed to drive and maintain convection.  Equation
(\ref{one}) is then equivalent to the statement that the residual
entropy $S'\equiv S-S_r$ is constant on a surface of constant $\Omega^2$.

On a strictly mathematical level, however, $S_r$ need not be a particular
entropy profile---in equation (\ref{one}), 
it evidently could be any function of $r$ at all.
Within the theory of BBLW, $S_r$ itself is never directly used in constructing
the rotation profile.  Instead,
it is $f(\Omega^2)$ that is important.  The arbitrainess of $S_r$
endows equation (\ref{one})
with a much greater generality than is obvious at first glance, a point
that we will develop in \S 2 of this work.

If the ansatz (\ref{one}) is used in the governing dynamical equation
(the azimuthal component of the vorticity equation), a self-contained
quasi-linear partial differential equation
emerges for $\Omega^2$ whose characteristics
are precisely the isorotation curves of region 1.   These contours
take the very simple mathematical form
\beq\label{two}
r^2\sin^2\theta =A - B/r
\eeq
where $A$ and $B$ are constant along a given curve\footnote{More
generally, $r^2\sin^2\theta = A + B\Phi(r)$, where $\Phi$ is the
gravitational potential (Balbus \& Weiss 2010).}; this basic
structure is in excellent
agreement with the helioseismology data (BBLW).  If the isorotation
curve passes through some colatitude angle $\theta_0$ at some radius 
$r=r_0$, then the explicit equation for the isorotation curves is
\beq\label{three}
\sin^2\theta = \left( r_0\over r\right)^2\left[\sin^2\theta_0 -{\beta}
\left( {r_0\over r} - 1\right)\right].
\eeq
The dimensionless parameter
$\beta$ is a number of order unity, which might be different for
different contours.  

It may be possible to develop this formalism to explain the
structure of the isorotation
contours even beyond the bulk interior of the SCZ.
Recently, Balbus \& Latter (2010, hereafter BL) argued that          
a tachocline structure resembling that of the Sun
emerges from the assumption of a simple $P_2$ spherical
harmonic vorticity source that is present inside of some transition
radius $r_T$, and is otherwise $r$-independent.  More precisely, if ${\cal
D}/{\cal D}r$ represents the radial derivative taken along a $\theta(r)$
characteristic given by equation (\ref{three}),
and if $C$ is a constant, then the equation
\beq\label{four}
{{\cal D}\Omega^2 \over{\cal D}r} =C P_2(\cos\theta)
\eeq
reproduces the SCZ isorotation contours for $C=0$, and is able
to closely describe
the tachocline isorotation contours for $C<0$.
The angular velocity remains continuous at the transition radius
$r=r_T$; its derivative along the characteristic is discontinuous.
In the bulk of the SCZ,
the TWE characteristics and the isorotation contours
are indistinguishable.   Within the tachocline,
they are very different curves.

This development is quite suggestive, hinting that an understanding
of the Sun's internal differential rotation may not require a detailed
dynamical knowledge of convective turbulence.  
Instead, the powerful constraint of thermal wind balance may be used
to determine the angular velocity profile
$\Omega(r,\theta)$ that is created by turbulent
transport.  At the same time, important questions arise
which motivate the current work.  In this paper, there
is perhaps some progress
toward elucidating some answers.   But much remains to be done, 
particularly for the subsurface outer layers.  

{\em Why is the match between mathematical solution and observations
so good in the bulk of the SCZ?}  There seems to be a curious mismatch
between the ability of the mathematical $r^2\sin^2\theta
=A-B/r$ curves to reproduce
the helioseismology data so strikingly, and the more qualitative physical
arguments that motivate the precise form of the thermal wind equation
that is used.  The theory seems to work too well.
We would not expect, for example, constant residual entropy
and constant angular velocity surfaces to match {\em exactly.}  We
would expect there to
be a {\em tendency} to follow one another, and the most solar-like numerical
simulations do show such a tendency (Miesch et al. 2006).  But assuming an
exact match leads to precise agreement between calculation
and helioseismology observations.  Is there something more going on here?

{\em Does thermal wind balance necessarily break down in the tachocline?}
BL did, in fact, interpret their equation this way, but
is this essential?  Is it possible to retain thermal wind balance with a 
different functional
relation between angular velocity and residual entropy that keeps the
goodness of fit intact?

{\em What is the physical explanation for the angular structure
of the tachocline isorotation contours?}
Both the helioseismic morphology and the rotational dynamics seem to be 
completely dominated by quadrupolar-like forcing.  Why?  For that matter,
how accurately do we know that the angular behavior of the vorticity stress?
Is it precisely quadrupolar?

{\em What is happening in the Sun's outer layer?}  Do turbulent stresses
become important (Miesch \& Hindman 2011)?  Can we guess the
form of a simple $\theta$-dependent forcing as was done for the tachocline?
The surface distortion of the isorotation contours does not
display $P_2$ symmetry, it is strictly monotonic from pole to equator.
The simplest possibility is $\sin^2\theta$ forcing.  How well does this work,
and what is its physical origin?

An outline of the paper is as follows.  In \S 2, we review the results
of earlier work (BBLW, BL) in which it
was shown that thermal wind balance and informal arguments 
were used to construct
precise mathematical solutions for the isorotation contours
in the bulk of the SCZ through the tachocline.   We then show    
that the all-important relationship between the residual
entropy and $\Omega^2$ must be true in a well-defined
and physically interesting 
asymptotic limit, independently of other dynamical complications.   
These ``complications,'' however, are likely to
broaden the range of validity
of the aymptotic limit.  
In \S 3, we construct
a global, thermal wind model for the Sun's differential rotation profile, 
and show that it compares favorably with the helioseismology data.  The outer
layers are treated on an ad hoc basis, with a deeper investigation of the issues
presented in an Appendix.  The change in the form of the isorotation 
contours apparent upon entering the 
tachocline does not indicate a breakdown of global thermal wind balance,
but instead a local functional dependence of the residual entropy on both angular
velocity and colatitude angle $\theta$.  In \S 4, we
present a physical discussion of our results, explore its limitations,
and suggest directions for future research.

\section {Thermal wind balance in the solar convection zone}              

\subsection{Pure rotation model: governing equation} 

Throughout this paper, we will work with both spherical $(r, \theta, \phi)$ 
and 
cylindrical $(R, \phi, z)$ coordinates.  As noted in the Introduction, $r$ and
$\theta$ are respectively the spherical radius and colatitude angle.  The quantity
$\phi$ is the azimuthal angle, while $R$ and $z$ are the cylindrical
radius and vertical Cartesian coordinate respectively.  

%
%

The governing dynamical equation of our analysis is the $\phi$ component of 
the vorticity equation.  For steady flows this is
\beq\label{dix}
(\rho \vv\bcdot\del)\left(\omega_\phi\over \rho  r\sin\theta\right)-
(\bb{\omega}\bcdot\del) \Omega = {1\over(r\sin\theta) \rho^2}(\del\rho\btimes\del P)\bcdot
\bb{e_\phi}
\eeq
or, since by mass conservation $\del\bcdot(\rho\vv)=0$,
\beq\label{onze}
\del\bcdot \left[ {\vv\omega_\phi\over R} - {\bb{\omega} \Omega}
\right] = {1\over R\rho}(\del\ln\rho\btimes\del P)\bcdot
\bb{e_\phi}
\eeq
where $\vv$ is the velocity, $\omega$ the vorticity,
$P$ the pressure, and $\rho$ is the density.   Equation (\ref{onze})
implicitly defines a vorticity flux within the divergence operator 
on its left side.  The vector $\bb{e_\phi}$ is of unit norm, in the 
azimuthal direction.

The left side of equation (\ref{dix}) 
contains two terms, the first corresponding to advection 
by the poloidal velocity components 
of the vortensity (divided by cylindrical radius),
the second to vorticity distortion by differential rotation. 
For a purely azimuthal velocity field, only
the second (vorticity-distorting) term is present. 
The leads to what is commonly referred to as the {\em thermal wind equation}
(Thompson et al. 2003): 
\beq\label{trez}
-\cos\theta\left(\dd\Omega^2\over \dd r\right)_\theta + {\sin\theta\over r}
\left(\dd\Omega^2\over
\dd \theta\right)_r  \equiv
- \left( {\dd\Omega^2\over \dd z}
\right)_R = {1\over(r\sin\theta) \rho}(\del\ln \rho\btimes\del P)\bcdot
\bb{e_\phi}.
\eeq
An important simplification of equation (\ref{trez}) 
is effected by replacing,
with impunity, $\ln\rho$ by $-(1/\gamma)\ln P \rho^{-\gamma}$, 
where $\gamma$ is the adiabatic index.  Then, 
in terms of the entropy variable 
$\sigma = \ln P\rho^{-\gamma}$
and gravitational field $g=-(1/\rho)(\dd P/\dd r)$, we have:
\beq\label{therm}
\left({\dd\Omega^2\over\dd r}\right)_\theta -{\tan\theta\over r}
\left({\dd\Omega^2\over \dd\theta}\right)_r =
 {g\over\gamma r^2 \sin\theta  \cos\theta}\left({\dd\sigma\over\dd \theta}
 \right)_r
 \eeq
On the right side, we have dropped the term proportional to 
$(\dd\sigma/\dd r)(\dd P/\dd\theta)$ in favor of
$(\dd\sigma/\dd\theta)(\dd P/\dd r)$. 
This is justified because, while the radial gradient
of $\sigma$ may well exceed its $\theta$ gradient in the SCZ
(region 1), because of entropy mixing by convection it does so by
a factor far smaller than the pressure radial gradient exceeds {\em its}
$\theta$ gradient.  We will adopt equation (\ref{therm}) as our governing
equation, though it must break down, of course, inside the radiative
zone (cf. \S 3).

\subsection {Establishing a functional relation between $\sigma$ and $\Omega^2$}

As discussed in the Introduction and in BBLW, 
the entropy variable $\sigma$ appears in equation (\ref{therm})
only in the form of its $\theta$ gradient.   This means that any function differing
from $\sigma$ by another function depending only upon $r$ would serve
just as well as $\sigma$ itself.  There is, in other words, a gauge invariance
in the problem\footnote{
A similar issue formally afflicts the angular velocity, which is insensitive
to an additive function of $R$.  This ``geostrophic degeneracy'' is
lifted by fitting our $\Omega$ solution to an explicit set of
isrotation surfaces provided by observations.}.  

This observation motivated BBLW to introduce the residual entropy $\sigma'$,
defined by
\beq\label{onebis}
\sigma'(r, \theta) = \sigma(r, \theta) - \sigma_r(r)
\eeq
where $\sigma_r$ is the minimal adverse radial entropy profile needed
to both drive and maintain ongoing convection.  In this manner, we choose our
gauge.  The idea is that
mixing of entropy would eliminate {\em residual} entropy gradients within 
a convective cell, but the resulting nearly constant
value of $\sigma'$ within one convective
cell need not be the same constant in another cell.
BBLW went on to argue that 
the presence of shear would favor the tendency for    
long-lived coherent convective structures
to lie also within constant $\Omega$
surfaces.   Under these conditions, constant $\sigma'$ and constant 
$\Omega$ surfaces would tend to coincide.  

The mathematical implementation of this is a powerful analytic constraint.               
If $\sigma'=f(\Omega^2)$,
where $f$ is a function to be determined, then
\beq\label{constr1}
\sigma(r, \theta) = f(\Omega^2) + \sigma_r(r)
\eeq
(cf.\ equation [\ref{one}]).  
In this view, the combination of convection and rotation in the SCZ
constrains the two-dimensional
entropy to be an {\it additive} function $\Omega^2$ and $r$.  

Let us take a somewhat more formal viewpoint for the moment. 
On purely mathematical grounds, for this development to be valid, 
all that is required is that 
the constraint embodied by equation (\ref{constr1}) be satisfied for 
whatever reason.
If it is, the ensuing thermal wind partial differential equation becomes
\beq\label{therm2}
\left({\dd\Omega^2\over\dd r}\right)_\theta -
\left[ {\tan\theta\over r}
 + {gf'(\Omega^2)\over\gamma r^2\rho \sin\theta  \cos\theta}\right]
 \left({\dd\Omega^2\over\dd \theta}
 \right)_r \equiv {{\cal D}\Omega^2\over {\cal D}r}  =   0,
 \eeq
where $f'(\Omega^2)$ is $df/d\Omega^2$, and we have made use of the ${\cal D}$
notation of equation (\ref{four}).  The solution to equation (\ref{therm2})
is that $\Omega$ is constant along the contours
given explicitly by equation (\ref{three}).  
It is remarkable that no
knowledge of $f$ is necessary to extract the basic functional form of the
contours: $R^2$ is just a linear function of $1/r$.   
Since these simple contours lie essentially 
on top of the helioseismology data (BBLW), 
the mathematical formalism
at the very least
constrains any deeper physical theory for the origin
of $\Omega(r,\theta)$.  Equation (\ref{constr1}) and thermal wind
balance are empirically true.   The ultimate underlying physical cause
remains to be settled, however:  
the physical argument laid out in BBLW may or may not be the correct one.  

\subsection{Another approach to the $\sigma'-\Omega^2$ relationship}

In addition to a likely dynamical linkage between $\sigma'$ and $\Omega$,
there is another simple feature of our problem that supports
equation (\ref{constr1}) : the SCZ is rather thin.

Imagine a Sun-like star with a slender outer convective zone.  In the case
of the SCZ itself, the central radius is about $r_c=0.87 r_\odot$ and our
analysis extends $\pm 0.1 r_\odot$, so the relative spread in the radial
domain $\Delta r/r_c$ is some $11.5\%$.  The residual entropy $\sigma'$
is a two-dimensional function of $r$ and $\theta$, but it could equally
well be locally regarded, by the implicit function theorem, 
as a function of  $r$
and $\Omega^2$ (the sign of $\Omega$ should not matter)\footnote{There
are of course some technical restrictions that accompany this remark. 
The variable $\Omega^2$ should be well-behaved, free of internal extrema or
saddle points over the domain
of interest; surfaces of constant $\Omega$ should not be spherical.
These are not concerns for our present application.}.  We shall assume
this is valid throughout the SCZ.
Our motivation
for choosing these two quantities as independent variables, as opposed to,
e.g.\ angular momentum and $r$, is clear: we wish to turn the thermal
wind equation into a simple mathematical constraint for the surfaces
of constant $\Omega$.  Had $\Omega^2$ not been one of our dependent
variables, the task of extracting the form of the constant isorotation
contours from the thermal wind equation would be much more difficult.
Much the same technique is familiar in thermodynamics, where the choice
of functions for independent variables is somewhat arbitrary, but often
dictated by what quantity is being held constant during the transformation
of interest.

If the dispersion in $\Omega^2$ and $r$ in the SCZ is such
that $\sigma'(\Omega^2, r)$ never leaves the bilinear regime,
then equation (\ref{constr1}) will be directly satisfied,
with gauge invariance allowing us to use either $\sigma$
or $\sigma'$.  But an expansion in $\Omega^2$ is not essential.
If we expand $\sigma'(\Omega^2, r)$ in a Taylor series about $r=r_c$, 
there obtains
\beq\label{twobis}
\sigma'(\Omega^2, r) \simeq \sigma'(\Omega^2, r_c)+
\left(1-{r_c\over r}\right)\left( {\dd \sigma'\over \dd\ln r}
\right)_\Omega  
 + {\rm ...} \eeq
The second term, which we shall regard as a small, is reduced relative
to the first by an explicit factor of $1-r/r_c\simeq 0.1$.  But, as we
argued in the Introduction, if in the dynamical regime of the SCZ there
is already some tendency for constant $\sigma'$ and $\Omega$ surfaces
to coincide, which is another way of saying that {\em any partial
derivative of $\sigma'$ taken at constant $\Omega$ will be small.}
The point is that ``small'' is by no means infinitesimal:
the first-order correction to retaining the leading term in equation
(\ref{twobis}) is quadratic, not linear in small quantities.  
Two $10\%$ effects would combine to yield a $1\%$ effect.  It is the
mutually reinforcement of the tendency of $\sigma'$ and
$\Omega$ surfaces to blend (even if it is not more than a trend), 
plus the narrow shell approximation $|r-r_c|\ll r_c$, that renders
the $\sigma'(r,\theta)=\sigma'(\Omega^2)$ ansatz more robust 
than might at first sight seem apparent.  The narrowness of the
shell means that residual entropy variations are more
beholden to changes in $\Omega$ than to
the spread in $r$, and the process of convection reinforces this 
trend by constraining the radial residual entropy gradients
(at constant $\Omega$).  The Sun happens
to be in a felicitous
theoretical regime from this point of view: it has both
an order unity convective Rossby number (Miesch \& Toomre 2009)
and a relatively narrow convective layer.  

Is it at all useful to consider the dependence $\sigma'(\Omega^2,\theta)$?
This seems counterproductive in the SCZ, where $\Omega$ already has a
dominant $\theta$ dependence, and where we have just advanced arguments to the
effect that $\sigma'$ should be considered a function of $\Omega^2$ alone.
But we can also turn this argument around.  In the tachocline,
$\Omega$ is clearly dominated by its $r$ dependence.  Therefore, from a
global viewpoint, adopting a $\sigma'(\Omega^2, \theta)$ 
dependence would allow
both the tachocline and the SCZ to be treated on the same footing, without
necessarily abandoning thermal wind balance.  In this formalism, $\sigma'$ goes
from a two-dimensional function in the tachocline $(\Omega^2, \theta)$
to a simpler one-dimensional function in the SCZ $(\Omega^2)$.  Such 
a dynamical structure would not be unknown in the study of rotating fluids: 
a classical two-dimensional Ekman boundary layer joining onto a
one-dimensional interior Taylor column exhibits the same feature.
In \S 3, this idea will be more fully developed.  

\subsection{Intuitively understanding the convection zone}

\subsubsection{Mathematical description}

The current approach has helped us to understand the answer to the first
question posed in the Introduction: why do informal arguments work so
well when our mathematical solutions are compared with observations?
It is enlightening to examine the mathematical form of our simple
SCZ solution with a view toward the helioseismology data.

Let us start with the explicit solution (\ref{three})
for an isorotation contour in the form
\beq
\sin^2\theta = {\sin^2\theta_0 + \beta\over y^2} -{\beta\over y^3}
\eeq
where $y=r/r_0$.  Then, 
expanding about some fiducial radius $y_c$,
\beq
\sin^2\theta  = \left( {\sin^2\theta_0 + \beta\over y_c^2} -{\beta\over y_c^3}
\right)
+(y-y_c) \left( {3\beta\over y_c^4} -{2\sin^2\theta_0 +2\beta\over y_c^3}
\right) + ...
\eeq
which implies the existence of a constant $\theta$ contour (a radial spoke) if
\beq
y_c = {1.5\beta\over \sin^2\theta_0 +\beta}
\eeq
lies within the convection zone.  The data show that in fact it does so
for $\theta_0\simeq 30^\circ$.  For other angles 
not too near the poles or equator, the
deviation from this class of contour is evidently not large. 
But neither is the alignment perfect: at larger $\theta_0$ the spoke
is canted more parallel to the $z$ axis, while at smaller $\theta_0$
the cant is more orthogonal to the vertical direction. 
All this can be seen both in the helioseismology data,
and the leading order term of a Taylor expansion of our simple solution.

\subsubsection{Dynamics}

What is special about $\Omega\simeq\Omega(\theta)$ that much
of the convection zone is to first order well-described by this class of contour?   
It is by no means a direct consequence of our $\sigma'=f(\Omega^2)$
ansatz, yet it is the most striking feature of the SCZ, and must
involve the dynamics of heat convection and rotation.  We defer 
an extended 
discussion of this important point to a subsequent paper 
(Schaan \& Balbus 2011), but the basic picture is easily grasped. 
The coupled dynamics of entropy and angular velocity
gradients are indeed at the heart
of matter.  The most
efficient route for heat transport is spherical convection,
radial motions being the most direct path
to the surface.  However, an inevitable secondary 
consequence of turbulent convection in an initially uniformly rotating
system is the production of differential
rotation, a sort of dynamical pollutant in this case.  
The presence of differential rotation
introduces its own inertial force, pushing fluid parcels away from 
spherically radial motions toward {\em cylindically}
radial motions.  For a displacement $\bb{\xi}$, this acceleration
is of the form $-\bb{\xi}\bcdot\del\Omega^2$.   
(Note that this is {\em not} the Coriolis force.)
If spherical radial displacements dominate,
this oblateness-inducing force is formally eliminated when 
$\Omega=\Omega(\theta)$.  
Radially convecting elements could then
not help but move in constant $\Omega$ 
surfaces.   This closes the circle of theoretical 
self-consistency: the gross pattern of 
differential rotation in the Sun looks the way it does
because in the presence of unavoidable build-up of differential rotation,  
$\Omega\simeq\Omega(\theta)$ would produce a pattern that would not interfere
with the most efficient form of heat transport.  
The top priority of the convective zone
would remain uncompromised.  
Moreover, we have in the previous section noted that
the description of the $\Omega$ curves is {\em not} precisely        
$\Omega(\theta)$; the curves are canted slightly poleward everywhere but
at the highest latitudes.   This slight poleward cant emerges, in fact, 
from the linear theory
of convective displacements (Schaan \& Balbus 2011) as well, and
$\bb{\xi}\bcdot\del\Omega^2$ remains small.

\section {Global solutions of the isorotation contours}

\subsection{Entering the tachocline}

As one crosses from the SCZ into the tachocline, the isorotation contours
change their character abruptly.  
It is possible that this is due to the increased importance of
additional dynamics beyond pure rotational motion, and the thermal
wind equation breaks down.   
On the other hand, there is little reason to assume that 
the precise $\sigma'-\Omega^2$ relation in the SCZ        
should hold here; the $r$ component of 
$\del\Omega$ becomes vastly more important in the tachocline.
Let us assume that thermal wind balance continues to hold
in the tachocline.  We will make use of the ansatz
$\sigma'=\sigma'(\Omega^2, \theta)$, which is particularly apt,
since $\Omega$ now carries a strong $r$-dependence.   
This pronounced dependence on radius is of course a consequence of 
the boundary condition enforcing uniform rotation
at the radiative zone interface.  Once this zone is breached,
the radial entropy gradient starts to rise, and the approximation
of neglecting the term proportional to $\dd s/\dd r$ in the 
baroclinic contribution to the thermal wind equation eventually breaks
down.  But by this point, $\Omega$ will be so close to uniform
rotation that the shift of the isorotation contours
is unlikely to be important.

We turn now to an investigation of 
thermal wind balance in the tachocline.  

\subsection {The ansatz $\simga' = \sigma'(\Omega^2, \theta)$}

The thermal wind equation acquires a very different mathematical
structure if $\sigma'$ is taken to be a function of $\Omega^2$
and $\theta$, as opposed to $\Omega^2$ and $r$.  We will
assume that this functional dependence is well-defined locally in $r$
for the tachocline region.
The, the equation formally becomes
\beq\label{twt}
{\dd\Omega^2\over \dd r} -\left[ {\tan\theta\over r} +
{g\over\gamma r^2\sin\theta \cos\theta} \left(\dd\sigma'\over
\dd \Omega^2\right)_\theta\right]{\dd\Omega^2\over \dd\theta}
=
{g\over \gamma r^2\sin\theta \cos\theta}
\left(\dd\sigma'\over \dd\theta\right)_{\Omega^2} .
\eeq
This should be contrasted with the case $\sigma'=\sigma'(\Omega^2, r)$,
for which the equation takes the form 
\beq\label{twt2}
{\dd\Omega^2\over \dd r} -\left[ {\tan\theta\over r} +
{g\over\gamma r^2\sin\theta \cos\theta} \left(\dd\sigma'\over
\dd \Omega^2\right)_r\right]{\dd\Omega^2\over \dd\theta}
=0.
\eeq
What is striking here is that equation (\ref{twt}) has precisely the
same mathematical form invoked 
by BL to account for the rotation pattern
in the tachocline.  (The computed rotation pattern fit the data well.)
But this earlier study viewed the right side of the equation
as arising from {\em external} vorticity stresses.  Here, we see that
this same equation could equally well arise without the need to appeal to external
dynamics, emerging instead within the context of thermal wind balance.  

Notice the important distinction between $\dd\sigma'/\dd\Omega^2$
at constant $\theta$ versus constant $r$.  At constant $\theta$,
the derivative is perfectly well behaved in the tachocline.  At
constant $r$, the derivative is ill-behaved: a significant
change in $\sigma'$ can be accompanied by virtually no change
in $\Omega$.  For this reason, equation (\ref{twt}) is to be
preferred.

Comparison of equations (\ref{twt}) and (\ref{twt2})
reveals another simple but important fact.
If $\sigma'$ is a function of $\Omega^2$
alone in the SCZ (and it certainly appears to be so), equation (\ref{twt})
is valid globally.   The partial $\theta$ derivative on the right
vanishes in the SCZ, and comes alive within the tachocline.  
At the very least, this is a useful organizational framework 
for understanding solar rotation.  The truly daunting problem of 
mastering the underlying dynamics of the tachocline can be decoupled from the
strictly mathematical problem of finding self-consistent solutions
to equation (\ref{twt}).  For the latter, we may appeal directly
to the observations, and to very general and simple dynamical constraints.
While not directly addressing the underlying physical
structure of the tachocline, this approach is not without interest:
a compelling fit would be strong evidence for the assumption of thermal
wind balance.

\subsection {Tachocline solution}                                  

In BL, the characteristics of the SCZ 
continued into the tachocline, and the entire
inhomogeneous term on the right side of the
governing equation was assumed to be proportional 
to a function of $\theta$ 
only, the spherical harmonic $P_2(\cos\theta)$.

In the current theoretical formalism, we have created somewhat tighter
mathematical constraints for ourselves.  
The SCZ characteristics can now extend into the tachocline
(which BL found produced good agreement with the data) only
if $(\dd\sigma'/\dd\Omega^2)_\theta$ is a function of $\Omega^2$
alone.  This means that $(\dd\sigma'/\dd\theta)_{\Omega^2}$
can be a function only of $\theta$.  This in turn means that
the right side inhomogeneous term of equation (\ref{twt}) has a steep
$g/r^2\sim 1/r^4$ radial dependence.   This radial dependence
was {\em not} present in BL, since the entire transition zone
(from $0.77 r_\odot$ to $0.65 r_\odot$ was viewed as 
a local boundary layer.  How does the additional radial dependence,
a new feature of the current model,
affect the shape of the tachocline isorotation contours?  

In the tachocline, equation (\ref{twt}) may be written
\beq\label{uni}
{{\cal D}\Omega^2\over {\cal D} r} = {g\over \gamma r^2 \sin\theta\cos\theta}
\left( \dd\sigma'\over \dd \theta\right)_{\Omega^2}.
\eeq
The $\theta$ dependence of the right side of the equation may be inferred
from helioseismology data as being approximately proportional to 
$-P_2(\cos\theta)$, or $\sin^2\theta-2/3$.  We will consider a slightly
more general form, $\sin^2\theta-\epsilon$, allowing $\epsilon$ to be determined
by the best fit.  (Note that the $\theta$ dependence of $\sigma'$ itself must
then involve a superposition of $P_4(\cos\theta)$ and $P_2(\cos\theta)$,
as found in the simulations of Miesch et al. [2006].)  Consider 
the equation
\beq\label{unii}
{{\cal D}\Omega^2\over {\cal D} r} = {C\over r^4}(\sin^2\theta-\epsilon) .
\eeq
The formal solution to equation (\ref{unii}) is that $\Omega^2 (r)$ is given 
along a characteristic curve by
\beq\label{vingt}
\Omega^2(r) = \Omega^2(r_1) + \int^r_{r_1} {C\over r^4}
[\sin^2\theta(r) - \epsilon]
\, dr
\eeq
with 
\beq\label{twen}
\sin^2\theta(r) = \left( r_0\over r\right)^2\left[\sin^2\theta_0 -{\beta}
\left( {r_0\over r} - 1\right)\right].
\eeq
The coordinates $r_0$ and $\theta_0$ represent the starting and
defining point of the characteristic (the effective surface), and $r_1$ is a fiducial
radius marking here
the starting point of the tachocline.   In general, we will use equation (\ref{vingt})
with $r<r_1$, reversing the integration limits and changing the sign.
The $\beta$ parameter in general will depend
upon $\theta_0$, though it is simplest to take $\beta$ to be a global constant,
and we will often do so.  
The radius $r_0$ will always lie outside of the tachocline on the constant-$\Omega$
portion of the characteristic.  We use the notation 
$r_1=r_T$ for start of the SCZ-tachocline boundary, and
we may take $\Omega^2(r_1)=\Omega_0^2(\sin^2\theta_0)$.  
Then, along a characteristic (\ref{twen}), the variation of angular velocity is given by
\beq\label{twen1}
\Omega^2(r) = \Omega_0^2(\sin^2\theta_0)+\epsilon C F_3 -Cr_0^2 
(\beta + \sin^2\theta_0)F_5 +C\beta r_0^3 F_6
\eeq
where
\beq
F_j = F_j(r, r_T)=\int^{r_T}_r {du\over u^{j-1}}=
 {1\over j}\left( {1\over r^j} - {1\over r_T^j} \right)
\eeq

To apply this with a minimum of mathematical overhead to an interesting case,
consider a linear form for $\Omega_0^2(\sin^2\theta_0)$,
\beq
\Omega_0^2(\sin^2\theta_0) = \Omega_1^2 + \Omega_2^2\sin^2\theta_0, 
\eeq
with $\Omega_1^2$ and $\Omega_2^2$ constants.  This may be regarded as a
simplified model of the Sun's surface.  

For this particular form of $\Omega_0^2(\sin^2\theta_0)$,
the solution for $\Omega^2(r, \theta)$
is found by eliminating $\sin^2\theta_0$ between equations (\ref{twen}) and 
(\ref{twen1}).  The equation for the isorotational
contours is then found to be
\beq\label{isor}
\sin^2\theta = \left(r_0\over r\right)^2
\left[
{\Delta - \epsilon c_1 r_0^3F_3 +\beta c_1 r_0^5F_5 -\beta c_1 r_0^6F_6\over
1 - c_1r_0^5F_5} +\beta \left(1-{r_0\over r}\right)
\right]
\eeq
where we define the dimensionless constants $\Delta$ and $c_1$ by
\beq
\Delta = {\Omega^2 -\Omega_1^2 \over \Omega_2^2}, 
\quad c_1= {C \over r_0^3 \Omega_2^2}
\eeq
Note that the contour-labeling collection of constants comprising the quantity
$\Delta$ is numerically the same as $\sin^2\theta_0$. 
By choosing $c_1$ and $\epsilon$ appropriately,
equation (\ref{isor}) represents a global solution for both the bulk of the SCZ
and the tachocline. 

\subsection{The outer layer}

In the outermost layer of the Sun, $r>0.96r_\odot$, the convection
is likely to be vigorous, with velocities comparable to the sound
speed.  In this zone, there is no reason to expect that a simple 
thermal wind balance
prevails in the vorticity equation, and good reason to suspect
that turbulent forcing is involved (Miesch \& Hindman 2011).  
For deriving the form of the isorotation contours however, it is 
useful to have a simple phenomenological description of this region.  
An inspection of the helioseismology data indicates a $\sin^2\theta$
rotational morphology in the subsurface layers.   We have 
accordingly found that setting     
\beq\label{ol}
{{\cal D}\Omega^2\over {\cal D}r} = {C_o \sin^2\theta\over r_\odot-r}
\eeq
works extremely well as a governing equation 
for the outer layer isorotation contours.  
The right side of this equation is a departure from thermal wind
balance, and may be the outcome of some combination of
convectively-induced Reynolds stresses,
meridional circulation, and magnetic braking.  
Here, as
earlier, ${\cal D}/{\cal D} r$ is the radial derivative along the
SCZ contour, and $C_o$ is a universal constant.  In the formalism
of Miesch \& Hindman (2011) the baroclinic term is dropped, whereas
here it is incorporated in the ${\cal D}/{\cal D} r$ operator.
The right side of equation (\ref{ol}) would then correspond to the
${\cal G}$ term of Miesch \& Hindman, divided by $R\cos\theta$.
Whatever the origin of the right side forcing of (\ref{ol}),
this equation 
may be solved for the constant $\beta$ case
by characteristic techniques entirely analogous to those used
for the tachocline solution (\ref{isor}).   In fact,
the equations are simpler,
since $r$ may be set equal to $r_\odot$ everywhere, save the expression
$r-r_\odot$.

An investigation of the form of equation (\ref{ol}) based on retaining
poloidal circulation terms in the vorticity conservation equation (\ref{dix})
is presented
in the Appendix.  We find that the $\sin^2\theta$ dependence can be
readily reproduced
in such a picture, but the singular $1/(r_\odot-r)$ behavior seems
to require additional dynamics.  It is possible that the
increasingly steep radial entropy gradient plays a role via
the as yet neglected baroclinic term proportional to the
product of ${\dd P/\dd\theta}$ and ${\dd \sigma/\dd r}$.

\subsection{Results}

\subsubsection {Constant and varying $\beta$}

The analytic solutions that extend our characteristic techniques
into the tachocline and SCZ boundaries
are rigorous, we reemphasize, only if the $\beta$ parameter is a universal constant.
Otherwise there is a nontrivial coupling between the characteristic equations for
$d\theta/dr$ and $d\Omega^2/dr$ in the boundary layers.  (But not, it must be stressed,
in the bulk of the SCZ, which may handled rigorously in the case of a varying $\beta$.) 
We will develop accurate numerical models for
this case in a later study, but for the present we will begin with constant $\beta$
models.  These work well enough (remarkably so given their simplicity)
to foster confidence in our fundamental approach.
We will then take our expressions for the constant $\beta$ solutions, and
in effect abuse
them by using the same expressions with a varying $\beta$ chosen
to match the contours in the bulk of the SCZ.
Ignoring the differential equation couplings this would introduce
with the tachocline is justified if such couplings
were small, here a marginal assumption.   But the effect of this simple expediency
on the boundary layer isorotation curves produces an extremely
striking match to the data (cf. Figure [5]).  
This strongly motivates pursuing a more rigorous
numerical solution of the governing partial differential equation,
a project we defer to a later publication.

\begin{figure} [h]
\centerline{\epsfig{file=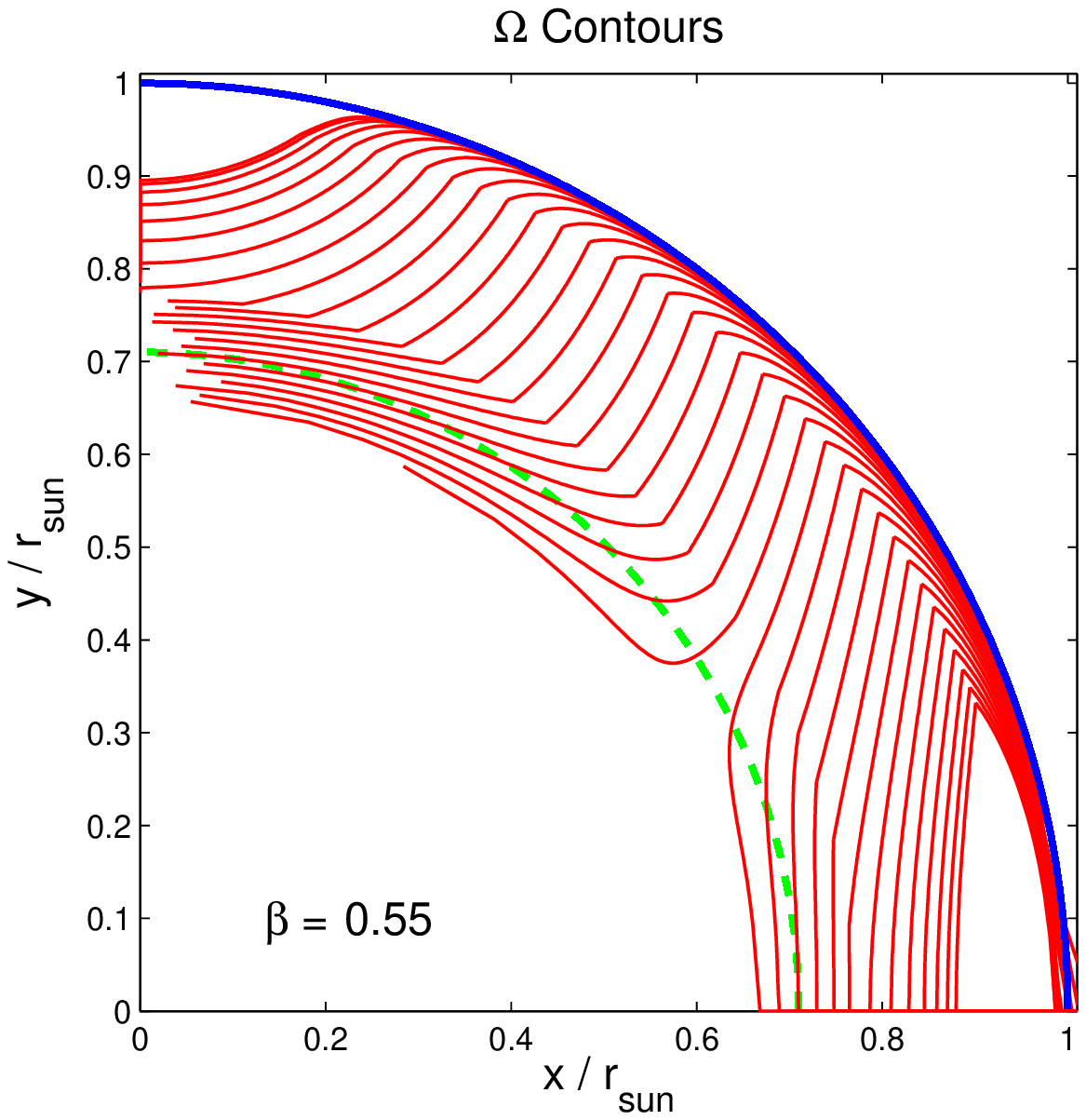, width=12 cm}}
\caption{The isorotation contours of the Sun according to equation
(\ref{isor}) with $\beta=0.55$, $r_0=0.896$, $r_1 = 0.96$, $r_T=0.77$ (solar radii),
$k_1= 1.745$. }
\centerline{\epsfig{file=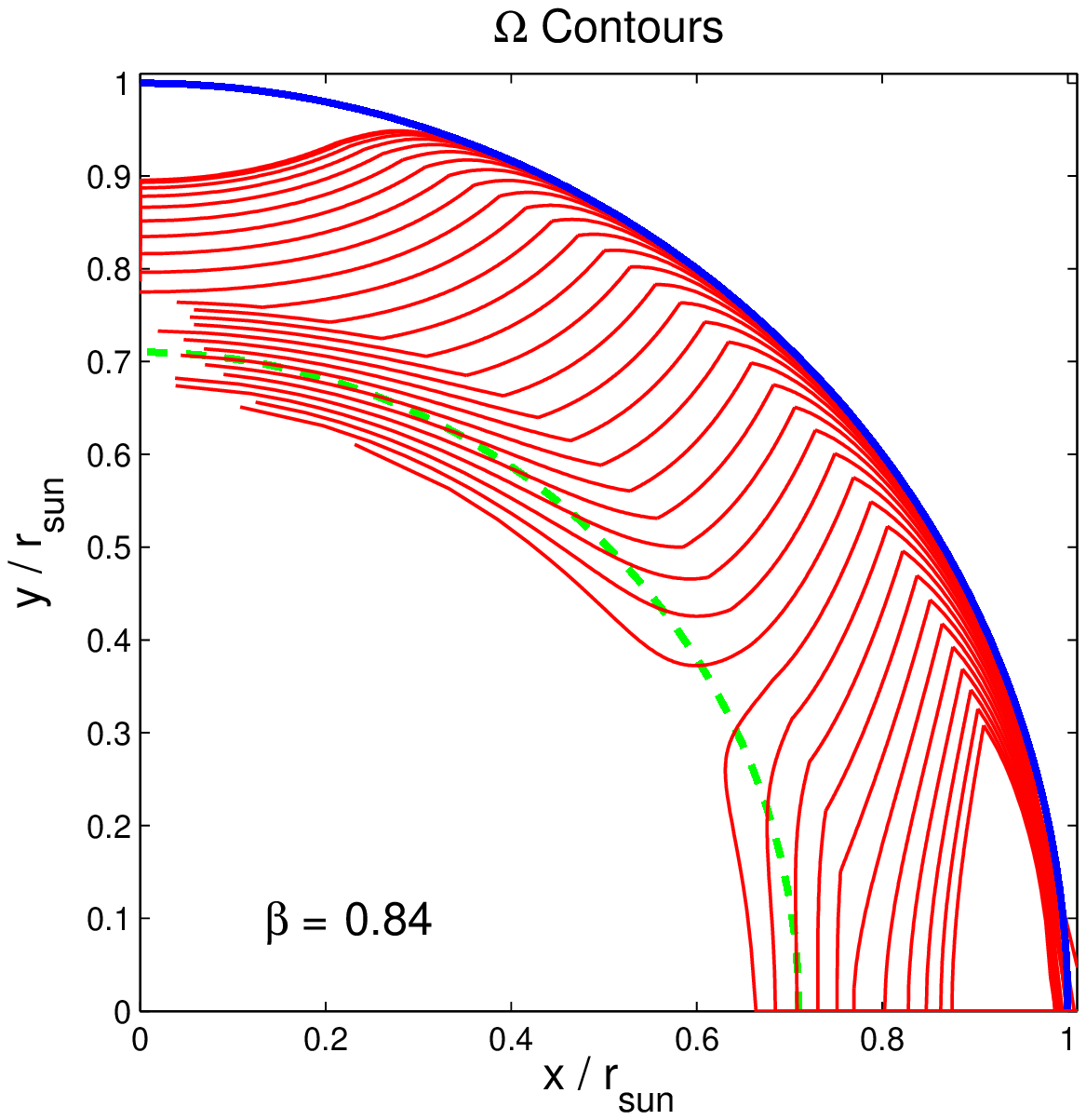, width=12 cm}}
\caption{As in Fig. [1] with $\beta=0.84$.}
\end{figure}
\begin{figure} [h]
\centerline{\epsfig{file=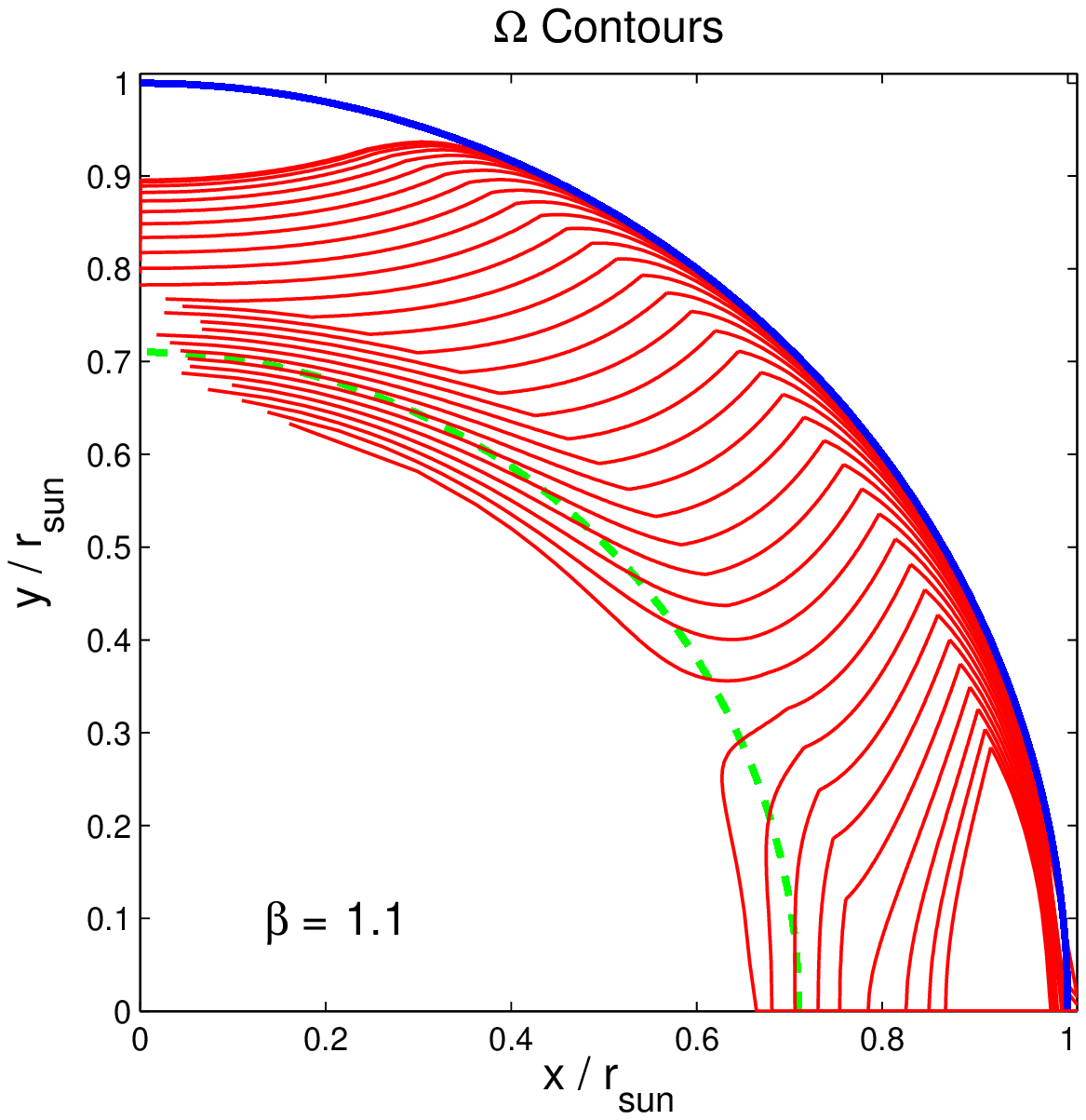, width=12 cm}}
\caption{As in Fig. [1] with $\beta=1.1$.}
\centerline{\epsfig{file=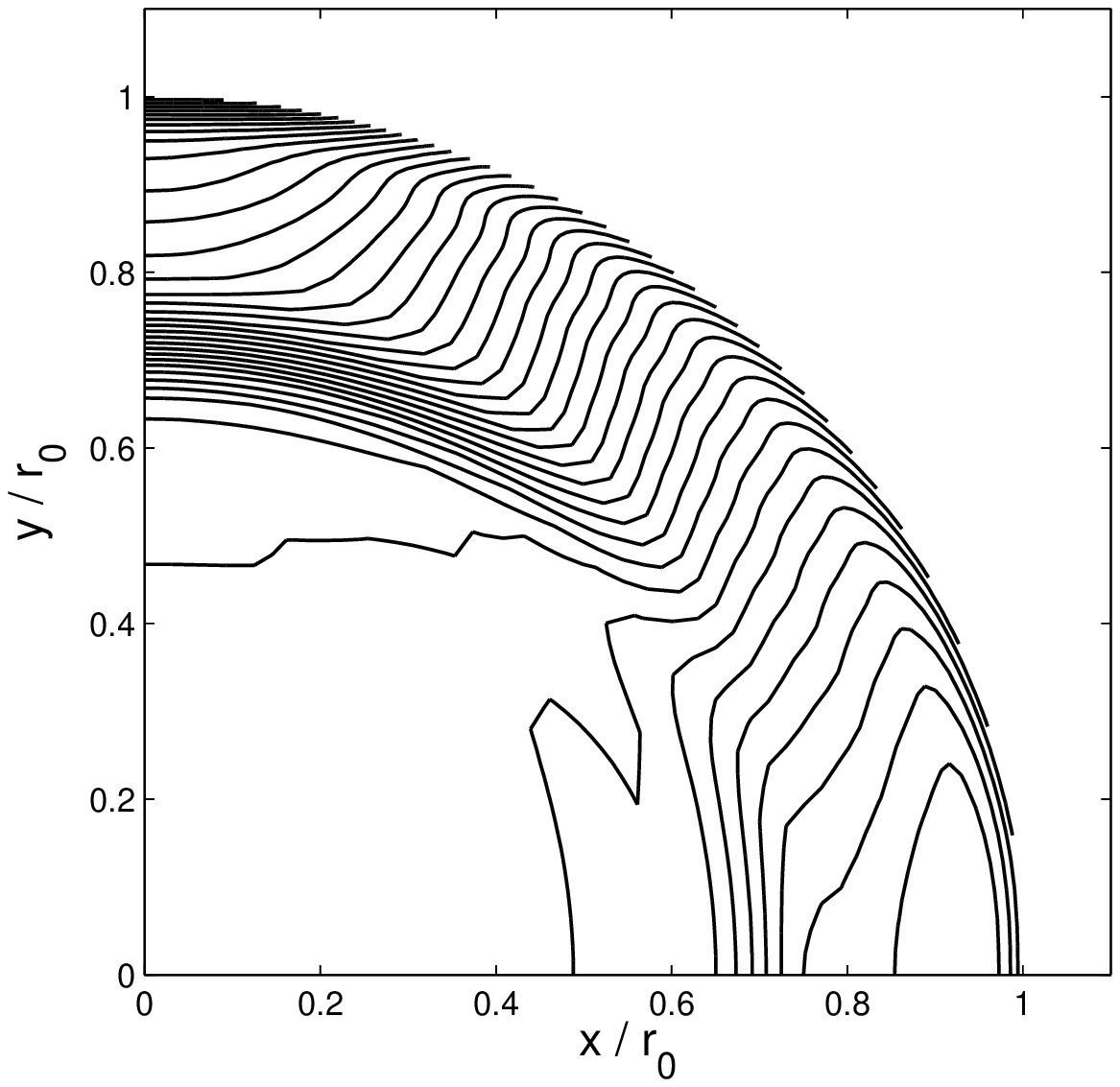, width=12 cm}}
\caption{Global Oscillation Network Group (GONG) isorotation contours 
(courtesy R. Howe).}
\end{figure}

Figures (1-3) show the constant $\beta$ isorotation contours 
(equation [\ref{isor}]) for
$\beta= 0.55$, 0.84, and 1.1 respectively.  The first value is chosen to 
match high latitude structure, the second to match the diverging
tachocline contours at $\theta\simeq60^\circ$, and the final $\beta$ value
to match the equatorial contour structure.  Overall, the global constant $\beta$
models clearly resemble the helioseismology data of figure 4.
In particular, the large $\beta$ contours near the bifurcation
point of the tachocline display a ``plateau'' followed by a precipitate,
cliff-like structure that is evident in the data.

In figure (5) we have overlayed the data with an analytic solution.
To effect this, we have used equation (\ref{isor}) but with
$\beta$ a function of $\sin\theta_0$:
\beq\label{bett}
\beta = 2.5\sin^2\theta_0 - 2.113\sin\theta_0 + .8205.
\eeq
The overlay shows an excellent, detailed, global match.

\begin{figure} [h]
\centerline{\epsfig{file=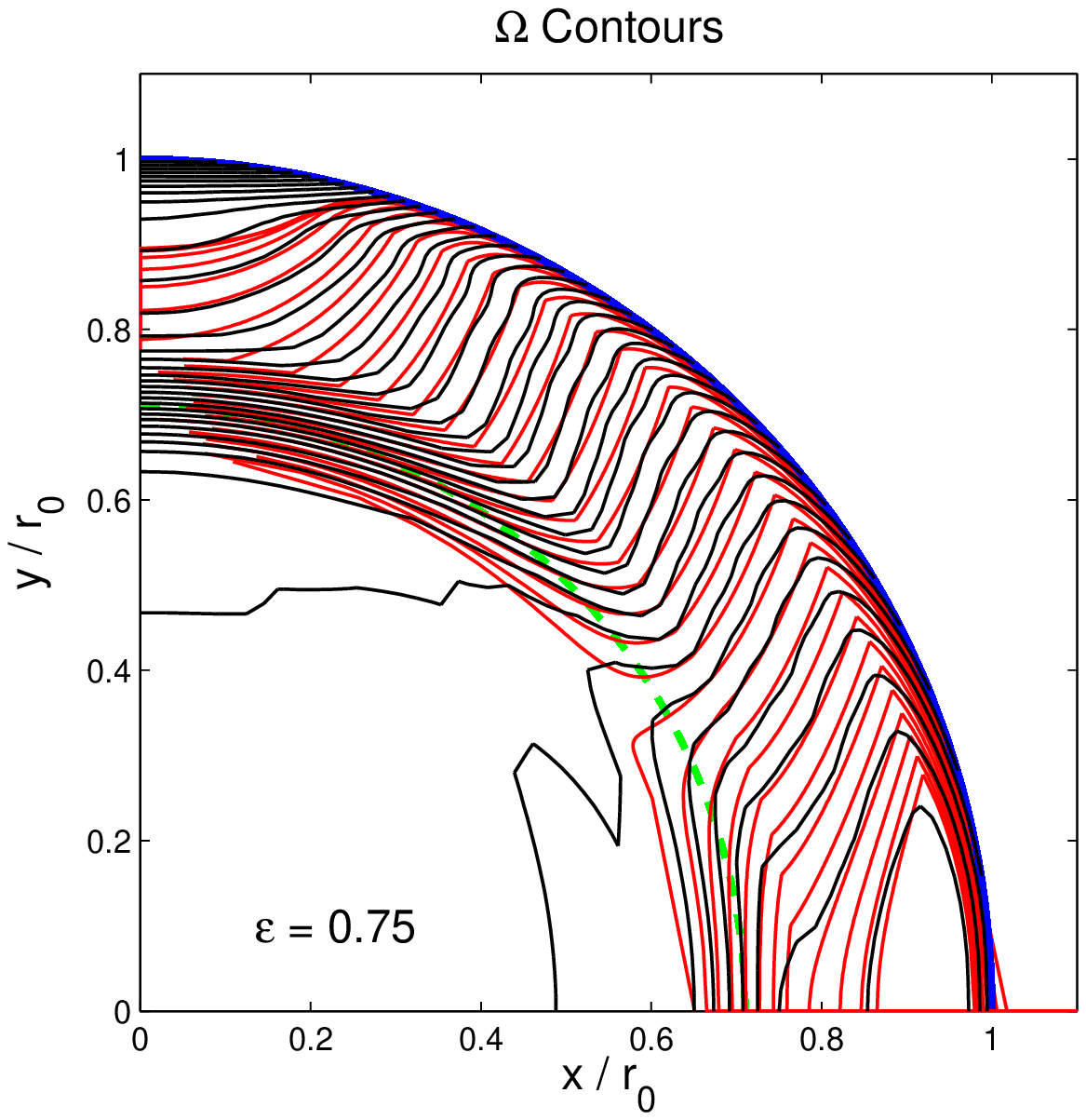, width=12 cm}}
\caption{GONG data (black) overlayed with fit from equations (\ref{isor})
and (\ref{bett}) (red).  $\epsilon=0.75$.  
Other parameters as in Fig. [1].
The contours
near the point of bifurcation at the radiative core
are very well-modeled.  
}
\end{figure}

\subsubsection{What is $\epsilon$?}

In figures (1-3), $\epsilon$ has been set
to $0.75$.  This parameter sets the location at which the
isorotation contours bifurcate in the tachocline.  What sets
its value?   

BL set $\epsilon$ equal to $2/3$ on the basis that this
fit the data rather well, and because $\sin^2\theta -2/3$ was
proportional to the spherical harmonic $P_2[\cos(\theta)]$,
a natural leading order structural modification to spherical 
symmetry.  But a closer investigation shows that there is 
a better data fit.  In figures (5) and (6) we show overlay plots
with $\epsilon=2/3$ and $4/5$ respectively.  While these 
fits are reasonable, they are clearly inferior to $\epsilon
=3/4$.  Is there something special about this value of $\epsilon$?

\begin{figure} [h]
\centerline{\epsfig{file=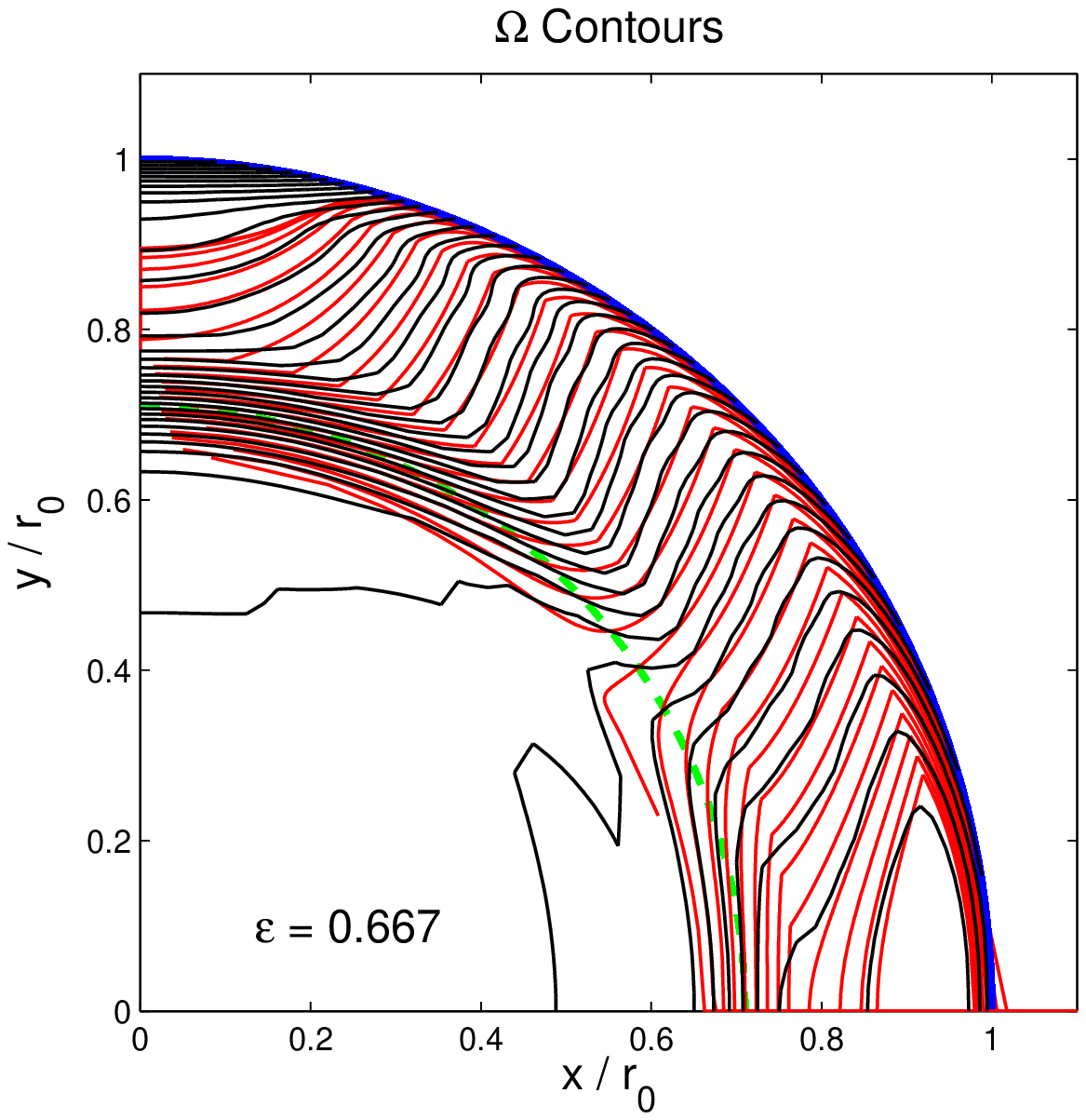, width=12 cm}}
\caption{GONG data (black) overlayed with fit from equations (\ref{isor})
and (\ref{bett}) (red).  $\epsilon=2/3$.  Other parameters as in Fig. [1].}
\end{figure}
\begin{figure} [h]
\centerline{\epsfig{file=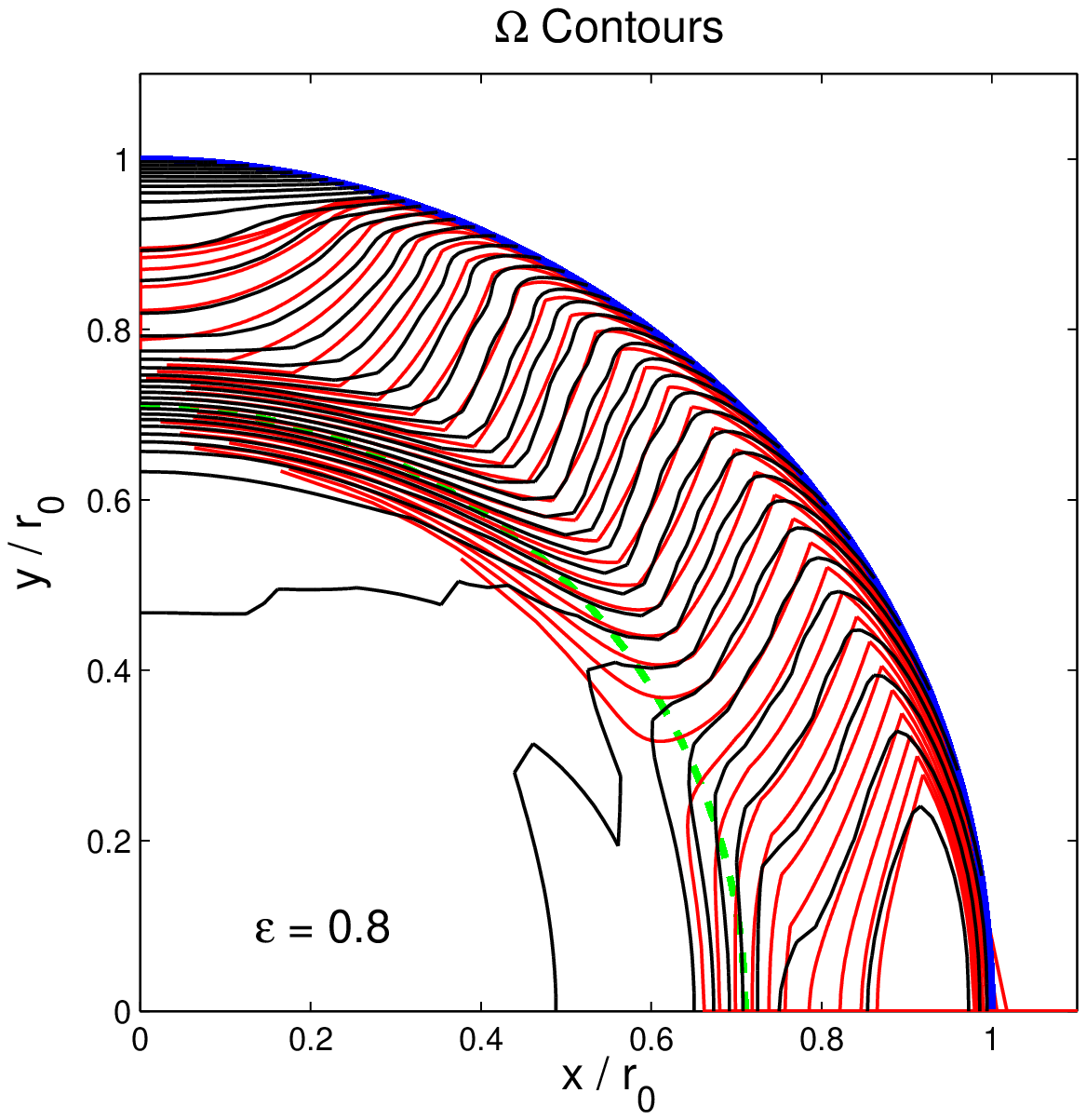, width=12 cm}}
\caption{GONG data (black) overlayed with fit from equations (\ref{isor})
and (\ref{bett}) (red).  $\epsilon=0.8$.  Other parameters as in Fig. [1].}
\end{figure}

One possible explanation is dynamical.  For example, what is the
torque exerted by our solution on the inner, uniformly rotating
radiative zone?  We assume that the couple is proportional to 
$R\dd\Omega/\dd r$, the $r\phi$ component of a viscous stress tensor.
We denote
\beq
\epsilon_n = {n+1\over n+2}
\eeq
so that $\epsilon_1$, $\epsilon_2$, and $\epsilon_3$ correspond to our
three choices.  Then the following elementary integral identity is satisfied:
\beq
\int_0^\pi \sin^n\theta\, (\sin^2\theta-\epsilon_n)\, d\theta = 0.
\eeq
The case $n=3$ is relevant to the calculation of the torque   
exerted by the tachocline on the inner uniformly rotating core
since this torque is directly proportional to the integral on
the left side of the equation. 
(One factor of $\sin\theta$
arises from the $r\phi$ stress tensor component, another from the 
lever arm $r\sin\theta$, and the third from the spherical area element.) 
In other words, when $\epsilon=\epsilon_3=0.8$, the torque exerted
by the tachocline on the interior vanishes. 
Since the couple to the core must act on time scales no shorter
than evolutionary, we might therefore expect $\epsilon=0.8$ to yield the best
fit.  

The best fit, however, is clearly $\epsilon=\epsilon_2=0.75$.  
This is close enough to 0.8 that there may be something correct about our 
argument,
but far enough off the mark that there is surely a piece of the puzzle missing.  
Moreover, it
really is the case that $\epsilon=3/4$ {\em is} the best fit
in our models, not just among the three values we display here.
While there is no compelling reason to believe that the value
of $\epsilon$ may be accurately deduced from the idealized 
assumption of a vanishing viscous torque, it is puzzling 
that an equally simple, but quite distinguishable, integral constraint
seems to fit the data so well.   It is as though we had mistakenly included an
additional factor of $\sin\theta$ inside our viscous torque integral.

For the present,
the explanation for why $\epsilon=\epsilon_2=0.75$ and not $
\epsilon_3=0.8$, corresponding
to a true vanishing stress at colatitude 
$\theta=60^\circ$ rather than 
$3.4^\circ$ further south, remains elusive.

\section {Summary}

In this paper, we have investigated global models of the angular rotation
profile of the solar convective zone (SCZ) using the toroidal component
of the vorticity equation and under the assumption of 
a dominant thermal wind balance.    
The model is global in that it
seeks to unite both the tachocline and the bulk of the SCZ within the thermal
wind formalism.  The subsurface outer layers have also been treated at a
quantitative, but more descriptive, level.  In effect, we have 
modeled the subsurface outer layers by an ad hoc source term to the 
thermal wind equation.                                  

The general nature of the isorotation surfaces in
the joint SCZ-tachocline structure seems to be a mathematical
consequence of a few simple principles, each noncontroversial
by itself:
i) thermal wind balance, as noted above;
ii) a narrow SCZ which may be treated
locally in $r$; iii) a tendency for surfaces of constant
angular velocity and residual entropy 
(see below) to blend; iv) an inner boundary condition of uniform
rotation at the tachocline base; v) vigorous radial transport in
the bulk of the SCZ; vi) a small couple between
the tachocline and the core.  When modeled by an idealized 
viscous prescription and the simplest possible
deiviation from spherical symmetry, assumption vi gives close, but not
perfect, agreement with helioseismology data.

As in earlier studies (e.g. BBLW, BL), we have introduced and
exploited the concept
of residual entropy, in essence 
the true entropy minus a suitable average over angles.
The residual entropy is, in fact, defined only up to an additive
function of $r$.  Numerical simulations are particularly
well suited to this concept: a Schwarzschild-unstable,
user-determined entropy profile $S(r)$ is imposed to drive the
requisite convective 
turbulence, and what one may call the ``response'' entropy 
$\tilde{S}(r,\theta)$ is then the 
quantity that is calculated (e.g. Miesch \& Toomre 2009 and 
references therein).  While $\tilde{S}$ may
formally differ
from the residual entropy $S'$ by compensating the radial
background $S(r)$, in practice it 
is not diffiult to extract a suitable
$S'$ from the numerical simulations.

The heart of our approach
is to regard the two-dimensional residual entropy as a function of
$\Omega^2$ and $\theta$ in the tachocline, but a function of $\Omega^2$
and $r$ in the bulk of the SCZ.  
In \S 2.3 we have argued that
the $r$ dependence of $\sigma'$ within the SCZ is ultimately unimportant.
In fact, it seems possible to arrive at the appropriate
form of the SCZ thermal wind equation for this latter case via
multiple routes;
the advantages of theoretical models that lead to the 
coincidence of constant $\sigma'$ and $\Omega$ surfaces have been discussed
elsewhere (Balbus 2009, BBLW, Schaan \& Balbus 2011).   
In any case, such behavior enjoys very clear numerical 
support (Miesch et al. 2006).

The explicit bulk-SCZ and tachocline solutions, equations (\ref{twen}) and (\ref{isor})
respectively, work reasonably well even in the constant $\beta$ limit, and
strikingly well when $\beta$ is permitted to vary.   Only the details
of the mathematical solutions, not their essential
structure, are sensitive to the choice of free parameters.  At this
point there seems little doubt that thermal wind balance is, 
in fact, the overall 
organizing principle for the tachocline and interior SCZ.  

We are, however, far from a complete solution.  
The mathematical equations contain free parameters, which must be
chosen to fit 
the helioseismology data.
We lack an understanding of how they are determined from first principles.
For example, the existence of a tachocline ``bifurcation point,'' 
where the isorotation
contours diverge either
toward the pole or equator, is clearly a consequence of the
$\theta$-independent boundary
condition imposed by the uniformly rotating core.   The location of this
bifurcation point depends upon our $\epsilon$ parameter, which reproduces
the data well if $\epsilon=0.75$, and much less well if $\epsilon$ changes
even by a little.  Why $\epsilon$ should be close to, but distinct
from, the value of 0.8 (corresponding to vanishing viscous torque on the
core) is not yet clear.  
Finally, the outer layers are not well understood.  Once again, 
the mathematical description of the isorotation contours seems simpler
than the underlying physics that gives rise to them.  A good match to the
helioseismology data is obtained from an approach
that uses thermal wind balance plus a simple $\sin^2\theta
/(r-r_\odot)$ external source.

We may thus end on a positive note.
If the work presented here is correct in its essentials, then the
beginnings of a deeper dynamical understanding of the rotation of the solar
convective zone are at hand.  It is gratifying that there is a certain
mathematical inevitability of the gross features of the
Sun's internal rotation pattern following from
thermal wind balance and a tight coupling between the 
rotation and the residual entropy isosurfaces.  
It is likewise encouraging that the more complex appearance of the isorotation
contours in the tachocline and outer layers seems to emerge from
relatively simple mathematics.  In short, there is no reason to think
that the dynamics of the solar rotation problem is intractable.  The blend
of numerical simulation and analysis promises to be a powerful combination
for elucidating the rotation profile of the convective regions of the Sun
and of other types of stars.  

\section*{Acknowledgements}  SAB is grateful 
to Princeton University Observatory for hospitality and
support as a Paczynskii Visitor, and 
to the Isaac Newton
Institute for Mathematical Sciences at Cambridge University, where this
work was begun.   We thank M. McIntyre and G. Vasil for their constructive
comments on an earlier version of this work, and the 
referee for a helpful, detailed report.  

\section*{Appendix: the outer layers}

The rotation pattern of the outer layers is
characterized by a robust $\sin^2\theta$ angular pattern, and 
the stress associated with this seems to depend on radius as
$1/(r-R_\odot)$.  We assume that there are
large scale poloidal velocity components in the outer layers,
arising perhaps from turbulent stresses present in this region
(Miesch \& Hindman 2011).
The introduction of nonuniform poloidal velocities $v_r$ and $v_\theta$ into
the flow causes the $\phi$ component of the vorticity, $\omega_\phi$,
to come into existence.  Because of our assumption of axisymmetry
($\dd/\dd\phi=0$) there are no changes at all to the poloidal components
of $\bb{\omega}$.  The addition of $v_r$ and $v_\theta$ velocities also
do not cause any cross terms between $\Omega$ and the poloidal components
in the azimuthal vorticity equation.  
This means that the poloidal components of the velocity, which 
advect changes in $\omega_\phi$, behave as ``external forcing'' with
regard to the angular velocity $\Omega$ in the governing 
partial differential equation.  

Start with the simplest possible circulation pattern consistent with
quadrupolar symmetry,
\beq
v_\theta= A(r)\sin\theta \cos\theta.
\eeq
Since mass conservation implies
$\del\bcdot(\rho\vv)=0$, the angular dependence of
$v_r$ must be the same as
\beq
{1\over r\sin\theta } {\dd (\sin\theta v_\theta) \over \dd \theta }
={A\over r} (2\cos^2\theta-\sin^2\theta) = {A\over r}(3\cos^2\theta - 1)
\eeq
In other words, $v_r$ has the angular dependence of $P_2(\cos\theta)$.
Letting
\beq
v_r = B(r) (3\cos^2\theta-1)
\eeq
we find that $B$ and $A$ are related to one another by the mass conservation equation
\beq
{1\over\rho }{d(\rho r^2B)\over dr} = - rA
\eeq

Next, we evaluate the terms in 
$\rho \vv\bcdot\del(\omega_\phi/\rho r\sin\theta)$ from the vorticity
equation (\ref{dix}).  
The terms involving the operator $v_r\dd/\dd r$ are
\beq
\rho v_r {\dd\ \over \dd r} \left[
{1\over\rho r^2\sin\theta}{\dd(rv_\theta)\over \dd r} -
{1\over\rho r^2\sin\theta}{\dd v_r\over \dd \theta} 
\right]=
\rho v_r \cos\theta {d\ \over d r} \left[
{1\over \rho r^2}{d(rA)\over d r} +{6B\over \rho r^2}
\right],
\eeq
which simplifies to
\beq
\rho B(3\cos^2\theta-1)\cos\theta {d\ \over d r} \left[
{1\over\rho r^2}{d(rA)\over d r} +{6B\over\rho r^2}
\right] 
\eeq
This leads to a forcing term that is directly proportional to $P_2(\cos\theta)$,
which is not what is observed.   By contrast,
the terms involving $v_\theta{\dd/\dd\theta}$ are
\beq
{v_\theta\over r^3}{\dd\ \over \dd\theta}\cos\theta\left[
{d(rA)\over d r} +6B\right]=
- {A\sin^2\theta \cos\theta \over r^3}\left[
{d(rA)\over d r} +6B\right]
\eeq
This gives a second forcing term proportional to $\sin^2\theta$,
which is consistent with the observations.  

Gathering the two terms together, 
\beq\label{XX}
\rho\bb{v}\bcdot\del\left(\omega_\phi\over \rho r \sin\theta\right) = 
\rho\cos\theta\left[
B(3\cos^2\theta -1 ){d\ \over d r} - {A\sin^2\theta\over r}\right]
\left[{1\over \rho r^2}{d(rA)\over d r} +{6B\over \rho r^2}\right]
\eeq
which, on the right side, is equal to 
\beq\label{p2sin2}
\rho\cos\theta\left[
B(3\cos^2\theta -1 ){d\ \over d r} + \sin^2\theta
{1\over \rho r^2}{d(\rho r^2 B)\over d r}\right]
\left[-{1\over \rho r^2}{d\over d r}{1\over\rho}{d\ \over dr}
(\rho r^2 B)+{6B\over \rho r^2}\right].
\eeq

Thus, there are two possible angular behaviors of the vorticity advection terms
that emerge from the analysis: $P_2(\cos\theta)$ and, possibly in 
superposition, $\sin^2\theta$.  
For the outer layers, however, we require forcing that
is proportional only to $\sin^2\theta$.  This means that the terms
in the second square bracket of equation (\ref{p2sin2}) must be a {\em constant,}
which we denote as $C$.  
Standard models for the outermost layers in the Sun 
seem to give something close to $\rho=\rho_0(1-x)^3$ for the
density, where $\rho_0$ is a characteristic density and
$x=r/R_\odot$, the radius 
in units of the solar radius.  
Then, with $u=1-x$,  $\rho=\rho_0 u^3$, this constancy
condition may be written as an inhomogeneous differential equation 
for $B$:
\beq
-{d\over d u}{1\over u^3}{d\ \over du}
( u^3 B)+6B= CR_\odot^2 \rho_0 u^3 .
\eeq
The solution for the homogeneous equation is 
\beq
B\ (homo) = c_1 {I_2(u\sqrt{6})\over u} +
c_2 {K_2 (u\sqrt{6})\over u}
\eeq
and the inhomogeneous solution is (to leading order in small $u$)
\beq
B\ (inhomo) = - {u^5\over 32} C\rho_0 R_\odot^2
\eeq

The inhomogeneous solution is well behaved near the solar surface
(i.e., for small $u$),
and the $I_2$ solution is
well-behaved at $u=0$, but the $K_2$ solution is singular.  
The mass flux, however, is finite.  Moreover, 
the disturbance should disappear as we move into the convective zone, so that
our solution must decrease with growing $u$.  This is not compatible with
$I_2$.  We therefore retain the $K_2$ solution---indeed, we are
looking for singular velocity behavior near the surface. 
Then, for small $u$ we have
\beq
A= -{1\over\rho} {d(\rho B)\over dx}= {1\over u^3} {d(u^3 B)\over du}
\eeq
or, 
\beq
A = {c_1\over u^3} {d(u^2 K_2(\sqrt{6} u))\over du} = {c_1\over u^2U}{d(U^2K_2(U))\over dU}
\eeq 
where $U=\sqrt{6} u$.  Using a standard Bessel function identity, this becomes
\beq
A= -{c_1\over u^2U}U^2K_1(U)= - {c_1U\over u^2}K_1(\sqrt{6}u)= 
-{c_1\sqrt{6} \over u}K_1(\sqrt{6}u)
\eeq
Therefore, to extract a solution with
$\sin^2\theta$ angular forcing in the outer layers
from the assumed circulation pattern,
the quantity $\omega_\phi/(\rho r
\sin\theta)$ and radial mass flux
must both be constant with depth, and the $\theta$ velocity
sharply decreasing with depth. 

If we now pull together the results of this mathematical excursion
and return to equation (\ref{XX}) to determine the nature of the forcing
we find i) it is rather straightforward to produce
$\sin^2\theta$ forcing; and ii) the sign of the forcing can be chosen by
a suitable choice of $C$ and $c_1$.  These are perhaps two important successes.
But in spite of the singular behavior of the $K_2$ function
near the surface, in this model the
vorticity stress is {\em not} 
singular.  The regular behavior is due to the appearance of the surface
density 
$\rho$ in the equation, which vanishes at the surface.
This is an important shortcoming of the model,
as the data 
seem to require logarithmic ``quasi-singular'' behavior
of the constant $\Omega$ contours near the surface.  
It may be that this feature can be captured by a more careful
consideration of the radially steepening entropy gradient,
which has been neglected here, lying outside the scope
of our analysis.

\end{document}